\documentclass[]{article}

\usepackage[utf8]{inputenc}
\usepackage{amsmath}
\usepackage{amssymb}
\usepackage{mathtools}
\usepackage{natbib}

\DeclarePairedDelimiter\ket{\lvert}{\rangle}

\DeclarePairedDelimiterX\braket[2]{\langle}{\rangle}{#1 \delimsize\vert #2}

\title{Derivation of A Relativistic Boltzmann Distribution}
\author{Alexander Taskov}

\begin{document}

\maketitle

\begin{abstract}
    A framework for relativistic thermodynamics and statistical physics is built by first exploiting the symmetries between energy and momentum in the derivation of the Boltzmann distribution, then using Einstein's energy-momentum relationship to derive a PDE for the partition function. It is shown that the extended Boltzmann distribution implies the existence of an inverse four-temperature, while the form of the partition function PDE implies the existence of a quantizable field theory of classical statistics, with hints of an associated gravity like gauge theory. An adaptation of the framework is then used to derive a thermodynamic certainty relationship.
\end{abstract}

\section{Introduction}
Relativistic thermodynamics is not a well understood subject. Much of its mainstream interest has died out since the beginning of the 20th century, having been overshadowed by the likes of quantum theory and general relativity. As such, there has hardly been any good consensus on matters as fundamental as the transformation law for temperatures (c.f. \cite{Einstein}, \cite{Ott}, \cite{Kibble}, \cite{Ekart}, or an introductory review by \citet{Review}).

Considering the statistical mechanical origins of quantum physics, as well as the relatively recent discovery of the thermodynamic nature of black holes, one might expect investigations into deeper thermodynamic theories to have taken more of a centre stage among physicists, although this has not been the case.

The purpose of this article will then be to derive, as straightforwardly as possible, a relativistic statistical mechanics and thermodynamics. The approach here will be essentially to exploit the symmetries between energy and momentum in relativity in the derivation of the canonical ensemble, and then take the results of this derivation to their furthest possible logical conclusion. From this we will show the existence of a four-temperature, as well as a quantizable field theory of statistical mechanics very much analogous to those of quantum field theory.

Throughout the article we take $c=k=\hbar=1$

\section{Relativistic Partition Function}
Consider an isolated, stationary ensemble of subsystems with conserved four momentum. Minimizing its entropy requires

\begin{equation}
\frac{\partial S}{\partial n_i} =  \frac{\partial}{\partial n_i} \log \frac{N!}{\prod_{i=1}^N n_i!} = 0
\end{equation}

where $N$ is the total number of subsystems and $n_i$ is the occupancy of the ith system state. The typical derivation (c.f. Boltzmann's original derivation translated by \cite{Boltzmann}) of the canonical ensemble would here apply the energy conservation restriction as a lagrange constraint, giving (after using stirlings approximation)
\begin{equation}
\log n_i + \beta \epsilon_i = 0 
\end{equation}
where $\beta$ is the inverse temperature (the lagrange multiplier) and $\epsilon_i$ is the energy of the ith subsystem state. This, however, reasonlessly favours the 0th component of the four-momentum. If we take into account all other conserved components of the four momentum, we get
\begin{equation}
\log n_i + \beta^0 p_i^0 - \beta^1 p_i^1 - \beta^2 p_i^2 - \beta^3 p_i^3 = 0 
\end{equation}
or, in Einstein notation
\begin{equation}
\log n_i + \beta^\mu p_i^\nu \eta _ {\mu \nu} = 0
\end{equation}
where the vector components $\beta^\mu$ are just a series of lagrange multipliers corresponding to the conserved components of the four-momentum, and $p^\mu_i$ are the components of the four-momentum of the ith subsystem state. This equation then implies
\begin{equation}
P_i = \frac{1}{Z}e^{-\beta^\mu p_i^\nu \eta _ {\mu \nu}}
\end{equation}
where $P_i$ is the probability of being in the ith system state, and $Z$ is the normalizing partition function.

There are two subtleties that must be addressed. Firstly, because the 3-momentum, unlike the energy, can be negative in the coordinate system, this relation appears to not only break isotropy, but also give a divergent normalization constant. This conundrum is solved by realizing that although the lagrange multipliers cannot vary with $p_i$ without becoming linearly independent of the vector $\nabla S$ in the space of $\{n_i\}$, they can take the form of a sign function, such that $P_i$ becomes both symmetric and normalizable, and the lagrange multipliers method only becomes invalid at a single point ($p_i=0$).

With this in mind, it is convenient now to take $p_i^\mu$ to be the magnitude of the four momentum components, such that the domain of $P_i$ is defined only on $(0,\infty)$ for all its dependent variables. This requires only thrice a doubling of the probability for any given 3-momentum, leaving a factor which we shall just absorb into the normalization.

The second problem is that the systems mass is also conserved, however this is not expressed in the relationship given here. The most straightforward way to include this would be as another lagrange constraint. There is, however, a much nicer, and more insightful way of doing this.	

Consider the partition function.
\begin{equation}
Z = \sum_{i=1}^\infty e^{-\beta^\mu p_i^\nu \eta _ {\mu \nu}}
\end{equation}
Now define

\begin{equation}
\partial_\mu = \frac{\partial}{\partial \beta^\mu}
\end{equation}

such that

\begin{equation}
\begin{split}
\partial_\mu \partial^\mu Z &= \sum_{i=1}^\infty p_{i \mu}p_i^\mu e^{-\beta^\mu p_i^\nu \eta _ {\mu \nu}}
\\
\\
&= \sum_{i=1}^\infty m_i^2 e^{-\beta^\mu p_i^\nu \eta _ {\mu \nu}}
\\\\
&=\langle m^2 \rangle Z
\\\\
&=m^2Z
\end{split}
\end{equation}

where we've defined $m^2$ to be the average subsystem mass squared. We now have the restriction on the partition function

\begin{equation}
(\partial_\mu \partial^\mu - m^2)Z = 0
\end{equation}

which encodes Einstein's mass energy relationship, and is of course very similar to the KGE

\begin{equation}
(\partial_\mu \partial^\mu + m^2)\psi = 0
\end{equation}

where here the partial derivatives are over position, rather than inverse temperature. The general solution to (10) is a linear sum of circular phases $e^{ipx}$. The general solution to (9), on the other hand, is not the partition function as given in (6), but in fact a linear sum of hyperbolic phases $e^{\sigma \beta p}$, where we define $\sigma ^2 = 1$.

It is straightforward to check that
\begin{equation}
e^{\sigma \beta p} = \cosh{\beta p} + \sigma \sinh{\beta p}
\end{equation}

and therefore represents a hyperbolic rotation. The general picture we then obtain is one where the statistics of quantum mechanics is governed by circular rotations, while the statistics of classical mechanics is governed by lorentz boosts. The gauge theory of some combination of these two statistics would then just be general relativity. The details of this are still unclear to the author, and this idea will be pursued only in later writings. First, let us further develop this relativistic thermodynamics.

\section{Four-Temperature}
Let us show that the inverse four-temperature is in fact a vector. We start from the derivation of the canonical ensemble for a system in contact with a thermal bath. Consider one of the subsystems of the previous section. If the four momentum forms a complete basis for the overall system, then for a subsystem of four momentum $p_i$ and total system momentum $p$ we have

\begin{equation}
S(p - p_i) \approx S(p) - \frac{\partial S(p)}{\partial p^\mu} p_i^\mu
\end{equation}

Using Boltzmann's formula,

\begin{equation}
P_i = \frac{1}{Z} \exp{ \Bigg( - \frac{\partial S(p)}{\partial p^\mu} p_i^\mu} \Bigg)
\end{equation}

Fundamentally, Equations (5) and (13) represent the same probability, and so we retrieve the relationship between entropy and the four-temperature

\begin{equation}
\beta_\mu = \frac{\partial S(p)}{\partial p^\mu}
\end{equation}

\begin{equation}
dS = \beta^\mu dp^\nu \eta_{\mu \nu}
\end{equation}

Note, although indices were used here quite liberally, they were used purely for notational purposes. No argument has yet relied on assumptions about the presence of a tensor.

Because entropy can be derived from just the microstates of the system, which by Galilean relativity alone should not change under a change of frame, entropy must be a lorentz invariant scalar. It is already known that the four-momentum is a four-vector, and therefore by (15) the four-temperature must also be a four-vector.

Although the zeroth component of the temperature is familiar, it is perhaps unclear what the 3-temperature represents. By the simple nature of its derivation, one can see it signifies the distribution of the 3-momentum occupancies. One can also note the similarity between (15) and the first law of thermodynamics, which implies that the $\beta_3 \cdot dp_3$ is somehow related to the work term $PdV$. If we take this equivalence literally (which we shouldn't, as the first law is purely classical) we will arrive at (for 1+1 dimensions)

\begin{equation}
\beta_3 = \beta^0 P \frac{dV}{dp_3}
\end{equation}

where $P$ is the pressure and $V$ the volume of the system. 

Finally, with equation (15) we can extend thermodynamics into general relativity to see that the presence of a gravitational field will have a direct effect upon the entropy. Once again, studies into gravity will be left to a later writing.

\section{Quantizing The Hyperbolic Partition Function}
In the first section we derived that the partition function must be representable as a linear combination of hyperbolic rotations by way of equation (9). It was already pointed out that this is in close analogy with quantum mechanics. Let us take this literally, and state explicitly the meaning of (9) in this context.

The modern interpretation of the KGE (10) is as describing and operator valued, quantizable field $\psi$. Adapting the typical quantization procedure requires us to consider the lagrangian and hamiltonian,

\begin{equation}
\mathcal{L} = \partial_\mu Z^\dagger \partial^\mu Z + m^2 Z^\dagger Z
\end{equation}

and therefore

\begin{equation}
H = \int (d\beta)^3 \hat{\pi}^2 + \nabla Z^\dagger \nabla Z - m^2 Z^\dagger Z 
\end{equation}

where

\begin{equation}
(e^{\sigma \eta})^\dagger = e^{-\sigma \eta}
\end{equation}

Note the limits of (18), which we will address when we come to its calculation, and see they will need some special treatment.

Carrying on with the quantization, it is easiest to first readapt Dirac's operator solution of the Harmonic Oscillator (c.f. \cite[pg.~136]{Dirac}) to a Harmonically decaying system, with the hamiltonian

\begin{equation}
\hat{H} = \frac{1}{2} \hat{p}^2 - \frac{1}{2} \omega^2 \hat{\beta}^2
\end{equation}

and then proceed to quantize the fields by simple analogy. Let

\begin{equation}
\begin{split}
b^\dagger = \sqrt{\frac{\omega}{2}} \hat{\beta} - \frac{\sigma}{\sqrt{2 \omega}} \hat{p}
\\
b =  \sqrt{\frac{\omega}{2}} \hat{\beta} + \frac{\sigma}{\sqrt{2 \omega}} \hat{p}
\end{split}
\end{equation}

Then because $[\hat{\beta},\hat{p}] = \sigma$ (Re: $k=1$) we get

\begin{equation}
\hat{H} = \omega \bigg(\frac{1}{2}- b^\dagger b\bigg)
\end{equation}

If we wish (18) to be quantized similarly, we require the field operators

\begin{equation}
\begin{split}
\hat{Z} &= \int_{-\infty}^\infty \frac{dp^3}{(i \sigma 2 \pi)^3} \frac{1}{\sqrt{2\omega_p}}
\bigg( b_p e^{\sigma \beta p} + c_p^\dagger e^{-\sigma \beta p} \bigg)
\\\\
\hat{Z}^\dagger &= \int_{-\infty}^\infty \frac{dp^3}{(i \sigma 2 \pi)^3} \frac{1}{\sqrt{2\omega_p}}
\bigg( b_p^\dagger e^{-\sigma \beta p} +  c_p e^{\sigma \beta p}   \bigg)
\end{split}
\end{equation}

\begin{equation}
\begin{split}
\hat{\pi} &= \int_{-\infty}^\infty \frac{dp^3}{(i \sigma 2 \pi)^3} \sigma \sqrt{\frac{\omega_p}{2}}
\bigg( b_p e^{\sigma \beta p} - c_p^\dagger e^{-\sigma \beta p} \bigg)
\\\\
\hat{\pi}^\dagger &= \int_{-\infty}^\infty \frac{dp^3}{(i \sigma 2 \pi)^3} (-\sigma) \sqrt{\frac{\omega_p}{2}}
\bigg( b_p^\dagger e^{-\sigma \beta p} - c_p e^{\sigma \beta p} \bigg)
\end{split}
\end{equation}

with the commutation relations
\begin{equation}
\begin{split}
[b_p,b_q]=[c_p,c_q]=[b_p,c_p]=[b_p^\dagger,c_p]=0
\\
[b_p,b_q^\dagger] = (2\pi)^3 \delta^{(3)}(p-q)
\\
[c_p,c_q^\dagger] = -(2\pi)^3 \delta^{(3)}(p-q)
\end{split}
\end{equation}

Now let us consider calculating the hamiltonian (18). Because (23) and (24) are essentially Laplace transforms, we require a special set of limits on the Hamiltonian.

\begin{equation}
\begin{split}
H &= \lim_{G \to \infty} (i \sigma 2 \pi)^3 \int_{-i \sigma G}^{i \sigma G} d\beta^3 \hat{\pi}^2 + \nabla Z^\dagger \nabla Z - m^2 Z^\dagger Z 
\end{split}
\end{equation}

where the complex coefficient is to ensure that the Hamiltonian is real. Substituting our relations and grinding through algebra retrieves

\begin{equation}
H= \int_{-\infty}^\infty dp^3 \omega_p \bigg( (2\pi)^3 \delta(0) - b_p^\dagger b_p - c_p^\dagger c_p \bigg)
\end{equation}
where we have used

\begin{equation}
i\sigma 2 \pi\delta(p-q) = \lim_{G \to \infty} \int_{-i \sigma G}^{i \sigma G} d\beta e^{\sigma \beta (p-q)}
\end{equation}

Interestingly, the hamiltonian (27) gives negative energies (already seen in the hamiltonian (20)) for the vacuum where the equivalent quantum mechanical calculation would give positive energies. This perhaps suggests that with infinitesimally differing limits on momentum, combining this field theory with quantum field theory would give a small but finite vacuum energy density. It was already mentioned that such a combination might create a gravity like gauge transformation, so combining these and retrieving a cosmological constant like quantity would be desirable. Alternativey, perhaps this quantization is unphysical, and either requires fermionic quantization, or is simply invalid entirely.  

\section{Probability Density and Statistical Operators}
Although interpreting the hyperbolic partition function as a field would be the theoretically closest analogy, its practicality is limited essentially as far as quantum field theory calculations. So, rather than delve into the field theory we shall for now simply take the old probability density approach.

The most obvious analogy to make is between the partition function and the wavefunction, with the eigenstates of momentum and temperature being the hyperbolic rotations $e^{\sigma \beta p}$. The inner product corresponding to the fourier transform of quantum mechanics is then

\begin{equation}
\begin{split}
\braket{p}{Z}& = \lim_{G \to \infty} \int_{-i \sigma G}^{i \sigma G} d\beta e^{-\sigma \beta p} Z(\beta)
\\
\braket{\beta}{Z} &= \frac{1}{i \sigma 2 \pi} \int_{-\infty}^{\infty} dp e^{-\sigma \beta p} \tilde{Z}(p)
\end{split}
\end{equation}
with the probability of a given momentum now being

\begin{equation}
P(p) = \tilde{Z}(p)^*\tilde{Z}(p)
\end{equation}

where we recall that $\sigma^* = -\sigma$
The operator corresponding to these eigenstates $\ket{p}$ is

\begin{equation}
\hat{p}^\mu = \sigma k \eta^{\mu \nu} \partial_\nu
\end{equation}

with the commutation relation

\begin{equation}
[\hat{\beta_\mu} , \hat{p_\nu}] = \sigma k \eta_{\mu\nu}
\end{equation}

This allows us to apply an operator calculus directly to the partition function, and so retrieve an uncertainty relationship between $\beta$ and $p$. Heisenberg's uncertainty relationship does not quite hold. Let us use a similar derivation to come to a new result.

Firstly note that because $(a+\sigma b)^*(a+\sigma b) = a^2 - b^2$ Shwarz' inequality becomes

\begin{equation}
u^2v^2 \leq  (u \cdot v)^2
\end{equation}

We now adapt the typical derivation of Heisenberg's uncertainty relation (c.f. \cite{Uncertainty}). Now let $\hat{Q}$ and $\hat{R}$ be hermitian operators such that the difference operators

\begin{equation}
\hat{Q}' = \hat{Q} - \langle Q \rangle
\end{equation}
are hermitian as well. These obey

\begin{equation}
|\braket{\hat{Q}^{'}}{Z}|^2 = \Delta q^2
\end{equation}

Two more hermitian operators can be formed

\begin{equation}
\begin{split}
\frac{1}{2} \{ \hat{Q}^{'} , \hat{R}^{'} \}
\\
\frac{\sigma}{2} [ \hat{Q}^{'} , \hat{R}^{'} ]
\end{split}
\end{equation}

allowing us to write

\begin{equation}
\hat{Q}^{'}\hat{R}^{'} = \frac{1}{2} \{ \hat{Q}^{'} , \hat{R}^{'} \} +
\sigma \frac{\sigma}{2} [ \hat{Q}^{'} , \hat{R}^{'} ]
\end{equation}

substituting these into the inequality (34) gives us

\begin{equation}
\begin{split}
\Delta q^2 \Delta r^2 &\leq  |\langle \hat{Q}^{'}\hat{R}^{'} \rangle |^2
\\
&\leq \bigg( \langle \frac{1}{2} \{ \hat{Q}^{'} , \hat{R}^{'} \} \rangle \bigg)^2 - \bigg( \langle\frac{\sigma}{2} [ \hat{Q}^{'} , \hat{R}^{'} ]\rangle \bigg)^2
\\
&\leq \bigg( \langle \frac{1}{2}\{ \hat{Q} , \hat{R} \}\rangle - \langle Q\rangle \langle R\rangle \bigg)^2 - \bigg(  \langle \frac{\sigma}{2} [ \hat{Q} , \hat{R} ] \rangle \bigg)^2
\end{split}
\end{equation}

which is, instead of an uncertainty relation, in fact a certainty relation. Unfortunately, the subtraction ensures the nonexistence of a simple, system independent form for (38). In spite of this, (38) in general tells us that there will never be information about the temperature without there being information about the energy, and vise versa. This is intuitive; in general, objects with high temperatures have with them associated high energies. This in contrast to momentum and position in quantum mechanics, where because momentum informs about changes in position and nothing more, one would expect information about momentum to be associated with a lack of information about position. Interestingly, these crucial differences arise solely from the different types of phases in the two theories, the same difference which arises between rotations and boosts in SR.

As a final note for this section, let us point out that dichotomy between $Z^*Z = a^2 - b^2$ and $\psi^*\psi = a^2 + b^2$ which hints that perhaps in that aforementioned gravity like gauge theory, probabilities are directly influenced by the metric of the space.
 
\section{Conclusion}
The basis of this article was the adjustment to the microcanonical derivation of the canonical ensemble to satisfy the relativistic symmetries associated with momentum and energy, and subsequently, the encoding of the mass energy relation in the differential equation (9). These ventures resulted in the definition of a four-temperature corresponding to the systems four-momentum distribution, as well as a hyperbolic, quantizable partition field. It was shown that the four temperature does indeed transform as a four-vector, and that all observables of the theory satisfy a certainty relation. This certainty relation ensures every measurable quantity of the theory is correlated to every other measurable quantity to some finite degree.

These relationships leave lots to be explored. Although much of this work has already been solved in the context of quantum theory, any old result rehashed into this new form will have to be reinterpreted in terms of temperatures, energies and partition functions. Remaining to be covered here: conserved charges of the lagrangian, gauge theories of the partition field, partition field interactions, Dirac like partition fields, temperature transformations in curved spacetimes, the nature of entropy, the nature of partition function "collapse", quantum statistical mechanics and potentially associated gravitational theories, implications for black hole thermodynamics (etc.).

Before any of the more exotic ideas are pursued, it would of course be desirable to draw both more concrete results from this which might be compared to experiment, as well as more complete interpretations. If the ideas presented here do indeed hold predictive power, they will not only be useful in studying statistical mechanics, but also for experimenters to perform studies on quantum mechanics, and in general come to a better understanding of the nature of probability and the meaning of observation. In the case they are not predictive, the mathematical symmetry is, nonetheless, interesting. 

\bibliography{RelativeThermo}
\bibliographystyle{unsrtnat}

\end{document}